# Mass Spectra and Regge Trajectories of Δ Baryons

Chandni Menapara*, Chetan Lodha and Ajay Kumar Rai

Department of Physics, Sardar Vallabhbhai National Institute of Technology, Surat-395007, Gujarat, India

*Email: chandni.menapara@gmail.com

**Abstract**

In contrast to past studies, the current paper is focused on baryons, and all four isospin states have been independently generated using u and d quarks with various constituent masses. The hypercentral Constituent Quark Model (hCQM) serves as the theoretical foundation for computing the resonance masses. The spin-dependent and first order correction terms are added to the confining potential, which is assumed to be in linear form. The resulting results have been contrasted with a wide range of methodologies and experimentally practicable states. Regge trajectories for (n, $M^2$) and (J, $M^2$) have also been displayed in addition to mass spectra.

1. Introduction

The hadrons, not just composite of valence quarks but their internal interaction leading to color confinement, gluons pose many questions to be addressed to reveal the degree of freedom responsible for the observed behaviour. Hadron spectroscopy has been aimed to obtain resonance masses of hadrons and explored through various approaches. the hadrons made by light quark flavours play an essential role in the interpretation of many reactions like in production processes of heavy quarks, heavy-ion collisions, etc. It is also of great interest to understand astrophysical systems like neutron stars. Also, this work is motivated in an attempt to understand all the light baryon properties particularly Δ baryon which comprises of four isospin partners formed with u and d quarks [1]. The experiments at Jefferson Lab, MAMI, ELSA, GRAAL, HADES-GSI [2] have been striving through the years to collect as much information about excited state of hadrons as possible. The upcoming facilities PANDA at FAIR-GSI shall be dedicated to light, strange baryons [3,4]. Light baryon has been summarized in some recent reviews [5,6].

Here, a non-relativistic quark model has been employed to obtain the mass spectra of all four isospin states of Δ by differentiating between u and d constituent quark mass. This is expected to aid in experiments as a large number of resonances are studied which can later be looked for through various decay channels. In addition, the Regge trajectories have been plotted to check for linear nature which shall be helpful in assigning the spin-parity to a given state.

## 2. Theoretical Framework: hCQM

As valence quarks dressed with gluons, quark-antiquark pairs, and all other interactions resulting in the total baryon mass, these constituent quarks are referred to as constituents. In this case, a non-relativistic, hypercentral constituent quark model (hCQM) has been used. The constituent quark mass for u and d quarks was assumed to be similar in our earlier work [7-10]. However, in the present work, the u and d constituent quarks masses have been modified respectively as $m_u = 290\ MeV$ and $m_d = 300\ MeV$ which allowed us to segregate the four isospin states of $\Delta$ baryon. The three-body interaction of quarks inside a baryon is described in the form of Jacobi coordinates $\rho$ and $\lambda$ which are obtained based on inter-quark distance $r_i$.

$$\boldsymbol{\rho} = \frac{1}{\sqrt{2}}(\mathbf{r_1} - \mathbf{r_2})\ ;\ \boldsymbol{\lambda} = \frac{1}{\sqrt{6}}(\mathbf{r_1} + \mathbf{r_2} - 2\mathbf{r_3}) \qquad (1)$$

The hypercentral Constituent Quark Model (hCQM) is reached through hyperradius x and hyperangle $\xi$.

$$x = \sqrt{\rho^2 + \lambda^2}\ ;\ \xi = arctan(\frac{\rho}{\lambda}) \qquad (2)$$

The potential to account for confinement and asymptotic freedom of quarks within a baryon is taken to be Coulomb-like part and a linear term as a confining part. As the model itself suggests, the potential is solely depended on the hyperradius x. It is noteworthy here that x indirectly is being contributed with the three-body interaction.

$$V(x) = -\frac{\tau}{x} + \alpha x \qquad (3)$$

To take into account the possible angular momentum quantum number J, spin-dependent terms are also added to the earlier potential terms.

$$V_{SD}(x) = V_{SS}(x)(\mathbf{S_\rho} \cdot \mathbf{S_\lambda}) + V_{\gamma S}(x)(\boldsymbol{\gamma} \cdot \mathbf{S}) + V_T \times [S^2 - \frac{3(\mathbf{S} \cdot \mathbf{x})(\mathbf{S} \cdot \mathbf{x})}{x^2}] \qquad (4)$$

Here, $V_{SS}(x)$, $V_{\gamma S}(x)$ and $V_T(x)$ are spin-spin, spin-orbit and tensor terms respectively.

In addition to above terms, a first order correction term with $\frac{1}{m}$ dependence has also been incorporated.

$$V^1(x) = -C_F C_A \frac{\alpha_s^2}{4x^2} \qquad (5)$$

where $C_F$ and $C_A$ are Casimir elements of fundamental and adjoint representation. The final Hamiltonian is as follows:

$$H = \frac{P^2}{2m} + V(x) + V_{SD}(x) + V^1(x) \qquad (6)$$

The Schrodinger equation with the hyper-radial part is numerically solved for calculating the excited state masses.

### 3. Results and Discussion

### 3.1 Mass Spectra

Using the above potential model, the masses are computed for 1S-5S, 1P-3P, 1D-2D, 1F states including few states from 1G, 1H and 1I which were not obtained earlier. The excited states are recalculated for all these isospin states of $\Delta$ baryon and compared with various results as shown in the table [1]. As the present work has attempted to separate the isospin states of all four $\Delta$ baryons, the ground state masses nearly vary by 2 MeV for each one but within the PDG range. A similar trend is observed in higher radial excited states of S-wave. The 2S mass predicted is very much near to the algebraic model.

The negative parity states have states ranging from 4 to 1 star status by PDG. It is observed from the table that with increase in J value, the predicted masses are under-predicted. However, not many models have different masses for every spin-parity state. In case of F-wave, $J = \frac{7}{2}$ state is the only known by PDG. The present masses are nearly 100 MeV below the range and hardly any comparison is obtained. Even with the small difference in the isospin state masses, this study is expected to aid in the decay channel studies as in the decay of heavy baryons, the final products are the light baryons. In addition, the light strange baryons play important role in other areas including astrophysics to understand the composition of celestial bodies. With these obtained masses, we have attempted to study the Regge trajectories as well as magnetic moments of $\Delta$ baryon. Here, a few models considered for comparison include: Bethe-Ansatz in U(7) model [11], relativistic interacting quark diquark model [12], semi-relativistic model [13], mass formula classification [15,16] and chiral quark model [17].

Table 1: Mass Spectra of all four isospin states of Δ in comparison with other models (in MeV).

| State | $J^P$ | $\Delta^{++}$ | $\Delta^+$ | $\Delta^0$ | $\Delta^-$ | [5] | PDG | Status | [11] | [12] | [13] | [14] | [15] | [16] | [17] |
|---|---|---|---|---|---|---|---|---|---|---|---|---|---|---|---|
| 1S | $\frac{3^+}{2}$ | 1228 | 1230 | 1232 | 1235 | 1232 | 1230-1234 | **** | 1245 | 1247 | 1231 | 1232 | 1232 | 1232 | 1232 |
| 2S | $\frac{3^+}{2}$ | 1603 | 1606 | 1610 | 1615 | 1611 | 1500-1640 | **** | 1609 | 1689 | 1658 | 1727 | 1625 | 1600 | 1659.1 |
| 3S | $\frac{3^+}{2}$ | 1922 | 1926 | 1932 | 1941 | 1934 | 1870-1970 | *** | | 2042 | 1914 | 1921 | 1935 | 1920 | 2090.2 |
| 4S | $\frac{3^+}{2}$ | 2241 | 2248 | 2257 | 2270 | 2256 | | | | | | | | | |
| 5S | $\frac{3^+}{2}$ | 2559 | 2570 | 2584 | 2602 | 2579 | | | | | | | | | |
| 1P | $\frac{1^-}{2}$ | 1618 | 1625 | 1630 | 1634 | 1625 | 1590-1630 | **** | 1711 | 1830 | 1737 | 1573 | 1645 | | 1667.2 |
| | $\frac{3^-}{2}$ | 1585 | 1591 | 1596 | 1603 | 1593 | 1690-1730 | **** | 1709 | 1830 | 1737 | 1573 | 1720 | | 1667.2 |
| | $\frac{5^-}{2}$ | 1542 | 1546 | 1552 | 1561 | 1550 | | | | | | | | | |
| 2P | $\frac{1^-}{2}$ | 1943 | 1944 | 1955 | 1965 | 1956 | 1840-1920 | *** | | 1910 | | 1910 | 1900 | | |
| | $\frac{3^-}{2}$ | 1907 | 1911 | 1921 | 1903 | 1919 | 1940-2060 | ** | | 1910 | | | 1940 | | |
| | $\frac{5^-}{2}$ | 859 | 1868 | 1874 | 1885 | 1871 | 1900-2000 | *** | | 1910 | 1908 | | 1945 | | |
| 3P | $\frac{1^-}{2}$ | 2262 | 2271 | 2280 | 2295 | 2280 | - | * | | | | | 2150 | | |
| | $\frac{3^-}{2}$ | 2226 | 2235 | 2246 | 2260 | 2242 | - | - | | | | | | | |
| | $\frac{5^-}{2}$ | 2179 | 2188 | 2199 | 2213 | 2193 | | | | | | | | | |
| 1D | $\frac{1^+}{2}$ | 1898 | 1898 | 1911 | 1919 | 1905 | 1850-1950 | **** | 1851 | 1827 | 1891 | 1953 | 1895 | 1910 | 1873.5 |
| | $\frac{3^+}{2}$ | 1860 | 1862 | 1873 | 1882 | 1868 | 1870-1970 | *** | 1936 | 2042 | 1914 | 1921 | 1935 | 1920 | 2090.2 |
| | $\frac{5^+}{2}$ | 1808 | 1814 | 1823 | 1832 | 1818 | 1855-1910 | **** | 1934 | 2042 | 1891 | 1901 | 1895 | 1905 | 1873.5 |
| | $\frac{7^+}{2}$ | 1744 | 1753 | 1760 | 1771 | 1756 | 1915-1950 | **** | 1932 | 2042 | 1891 | 1955 | 1950 | 1950 | 1873.5 |
| 2D | $\frac{1^+}{2}$ | 2218 | 2221 | 2234 | 2247 | 2227 | | | | | | | | | |
| | $\frac{3^+}{2}$ | 2179 | 2184 | 2196 | 2210 | 2190 | | | | | | | | | |
| | $\frac{5^+}{2}$ | 2127 | 2135 | 2146 | 2160 | 2140 | 2015 | ** | | | | | 2200 | | |
| | $\frac{7^+}{2}$ | 2062 | 2073 | 2084 | 2098 | 2078 | | | | | | | | | |
| 1F | $\frac{3^-}{2}$ | 2146 | 2153 | 2160 | 2181 | 2165 | | | | | | | | | |
| | $\frac{5^-}{2}$ | 2092 | 2099 | 2108 | 2126 | 2108 | | | | | | | | | |
| | $\frac{7^-}{2}$ | 2024 | 2033 | 2043 | 2058 | 2037 | 2150-2250 | *** | | | | | | | |
| | $\frac{9^-}{2}$ | 1942 | 1953 | 1966 | 1975 | 1952 | | | | | | | | | |
| 1G | $\frac{11^+}{2}$ | 2132 | 2145 | 2162 | 2178 | | 2300-2500 | **** | | | | | | | |
| 1H | $\frac{13^-}{2}$ | 2326 | 2339 | 2362 | 2379 | | 2794 | ** | | | | | | | |
| 1I | $\frac{15^+}{2}$ | 2512 | 2529 | 2554 | 2581 | | 2990 | ** | | | | | | | |

### 3.2 Regge Trajectory

The linear relation of total angular momentum quantum number J as well as principal quantum number n with the square of resonance mass $M^2$ is at the base. The plotting of all the natural and unnatural parity states allows us to locate if a given state is in accordance with the assigned $J^P$ value.

$$J = aM^2 + a_0 \quad (7)$$

$$n = bM^2 + b_0 \quad (8)$$

So, here figures 1 to 6 depicts the Regge trajectories for the isospin partners. It is noteworthy that all the calculated points fit well on the linear curve and are non-intersecting.

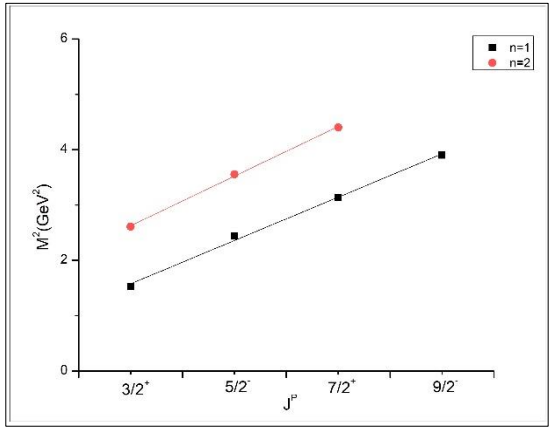

Fig 1: Regge trajectory $J \rightarrow M^2$ for $\Delta^{++}$

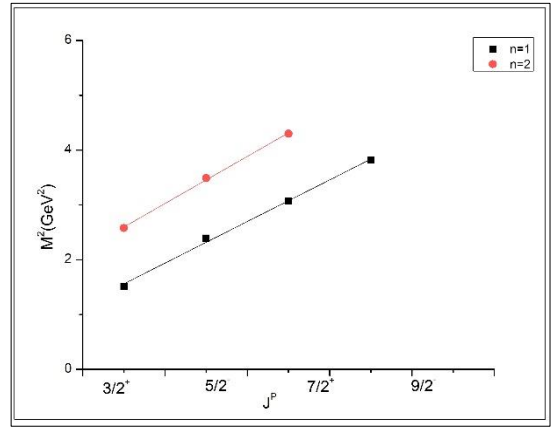

Fig 2: Regge trajectory $J \rightarrow M^2$ for $\Delta^+$

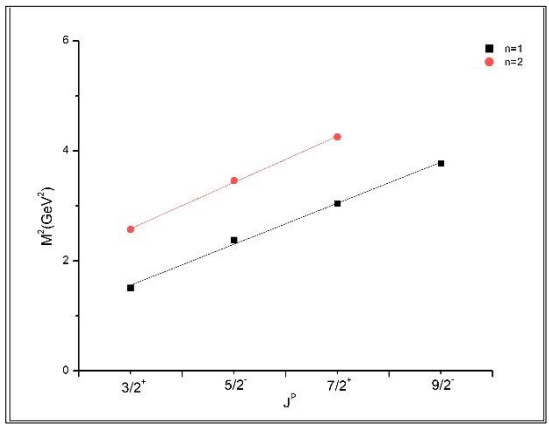

Fig 3: Regge trajectory $J \rightarrow M^2$ for $\Delta^-$

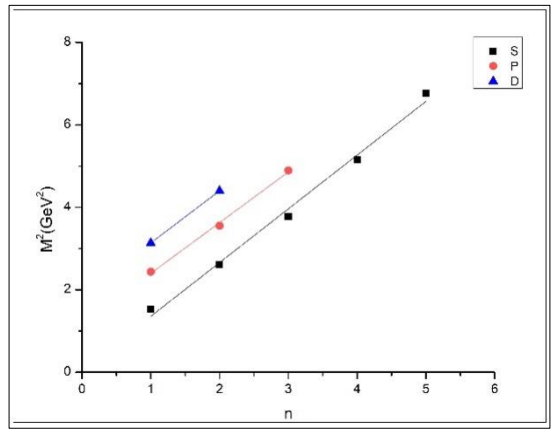

Fig 4: Regge trajectory $n \rightarrow M^2$ for $\Delta^{++}$

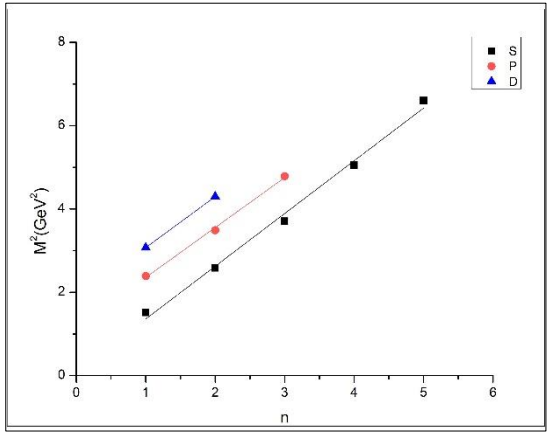 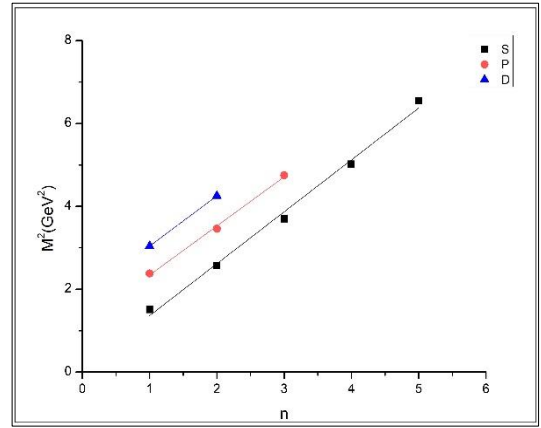

Fig 5: Regge trajectory n → $M^2$ for $\Delta^+$     Fig 6: Regge trajectory n → $M^2$ for $\Delta^-$

## 4. Conclusion

The idea of separately exploring the isospin states of $\Delta$ baryon has been implemented in the present work which is a modification to earlier calculated masses. The constituent quark masses for u and d quarks have been treated differently to obtain the radial and orbital excited state masses using the non-relativistic hypercentral Constituent Quark Model (hCQM). The potential incorporated consists of linear confining term, spin-dependent terms and first order correction terms.

The results have been compared with those of Particle Data Group (PDG) and with other approaches discussed above. The comparison has shown that low-lying states are well in accordance with experimental range. However, the higher $J^P$ value states for a given principal quantum number under-predicts as compared to PDG. The results are expected to aid in future experiments to determine missing resonances and in various decay channels.

The Regge trajectories have been plotted for (n,$M^2$) and (J,$M^2$) which are following the linear nature. This allows us to identify a given state for its spin-parity assignment on the Regge line. Differentiating the isospin resonance masses shall be important to study strong, weak and electromagnetic decay channels. This work is expected to support the upcoming experimental facilities at PANDA.

## 5. Acknowledgements

Gandhinagar for providing with an opportunity to present our work. Also, Ms. Chandni Menapara would like to acknowledge the support from the Department of Science and Technology (DST) under INSPIRE-FELLOWSHIP scheme for pursuing this work.